\documentclass[useAMS,usenatbib]{mn2e}
\usepackage{graphicx, rotating}
\usepackage{aas_macros}  
\usepackage{graphicx}
\usepackage{amssymb}
\usepackage{rotating}
\usepackage{multirow}
\usepackage[breaklinks,colorlinks=true,citecolor=black,linkcolor=black,urlcolor=blue,bookmarks=true]{hyperref}

\newcommand{\kms}{km\,s$^{-1}$}

\title[ SFH of And\,XVIII dSph galaxy]{Star formation history of And XVIII: 
a dwarf spheroidal galaxy in isolation
\thanks{Based on observations made with the NASA/ESA Hubble
Space Telescope, program GO-13442, with data archive at the Space
Telescope Science Institute. STScI is operated by the Association of 
Universities for Research in Astronomy, Inc. under NASA
contract NAS 5-26555.}
}
\author[L. N. Makarova et al.]{
L. N. Makarova$^{1}$\thanks{E-mail: lidia@sao.ru}, 
D. I. Makarov$^{1}$,
I. D. Karachentsev$^{1}$,
R. B. Tully$^{2}$,
L. Rizzi$^{3}$\\
$^{1}$Special Astrophysical Observatory, Nizhniy Arkhyz, 
Karachai-Cherkessia 369167, Russia\\
$^{2}$Institute for Astronomy, University of Hawaii, 2680 Woodlawn Drive, 
HI 96822, USA\\
$^{3}$W. M. Keck Observatory, 65-1120 Mamalahoa Hwy, Kamuela, HI 96743, USA
}
\begin{document}

\date{Accepted XXX. Received XXX; in original form XXX}

\pagerange{\pageref{firstpage}--\pageref{lastpage}} \pubyear{XXX}

\maketitle

\label{firstpage}

\begin{abstract}
We present a photometric study of the Andromeda\,XVIII dwarf spheroidal galaxy 
associated with M31, and situated well outside of 
the virial radius of the M31 halo. The galaxy was resolved into stars with 
Hubble Space Telescope/Advanced Camera for Surveys revealing the old red giant 
branch and red clump. With the new observational data we determined the Andromeda\,XVIII 
distance to be $D = 1.33_{-0.09}^{+0.06}$ Mpc using the tip of red giant branch method. 
Thus, the dwarf is situated at the distance of 579 kpc from M31.
We model the star formation history of Andromeda\,XVIII from the stellar photometry 
and Padova theoretical stellar isochrones.  
An ancient burst of star formation occurred 12--14 Gyr ago. 
There is no sign of recent/ongoing star formation in the last 1.5 Gyr.
The mass fractions of the ancient and intermediate 
age stars are 34 and 66 per cent, respectively, and
the total stellar mass is $4.2\times$10$^6\,M_{\odot}$.
It is probable that the galaxy has not experienced an interaction with M31 in the past.
We also discuss star formation processes of dSphs KKR\,25,
KKs\,03, as well as dTr KK\,258. Their star formation histories were uniformly 
measured by us from HST/ACS observations. All the galaxies are situated well 
beyond the Local Group and the two dSphs KKR\,25 and KKs\,03 are 
extremely isolated. Evidently, the evolution of 
these objects has proceeded without influence of neighbours.
\end{abstract}


\section{Introduction}
In recent years, the galaxies in the immediate vicinity of the Local Group have been the focus 
of intense and wide-ranging research \citep{mquinn, collins15, cr},
as these galaxies represent a unique laboratory for studies of the history of star 
formation in the Universe, the properties of dark matter 
and tests of the modern cosmological LCDM paradigm.
In particular, it is obvious that the Local Group can be especially promising in the search for
new dwarf galaxies. This search is especially important, bearing in mind the well known
"lost satellites" problem in cosmology, because the LCDM theory predicts more satellites
of giant galaxies than are found in reality. Recently we have detected and studied
several unusual nearby dwarf galaxies. These spheroidal objects are rather isolated,
or at the outskirts of the Local Group \citep{mak12, kar14, kar15}. 
Although the sample of galaxies is still small, 
they represent a class of dwarf galaxies that can help understand the problems 
of the formation and evolution of dwarf galaxies and groups of galaxies within the LCDM paradigm.

\begin{figure*}
\includegraphics[width=11cm]{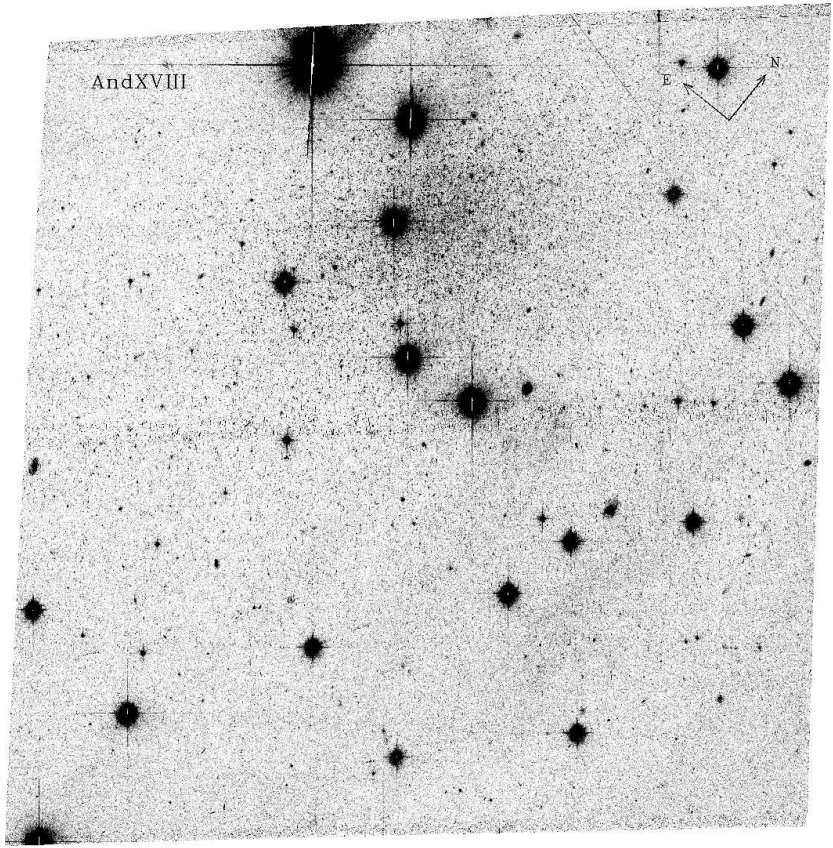}
\caption{\textit{HST}/ACS combined distortion-corrected  mosaic 
image of And\,XVIII in the \textit{F606W} filter. 
The image size is $3.4\times3.4$ arcmin.
}
\label{fig:ima}
\end{figure*}

The Local Group dwarf galaxy Andromeda\,XVIII is one more object in the sample of quite isolated
dwarf spheroidals. It was discovered by \citet{mccon2008} 
as a part of their CFHT/MegaPrime photometric survey of M31. The observations 
were made in {\it g} and {\it i} bands, and the galaxy was seen as a prominent 
overdensity of stars. The colour-magnitude diagram of And\,XVIII was obtained 
by \citet{mccon2008}, and red giant branch stars were clearly visible.
\citet{tollerud} observed spectra of a number of red giant stars belonging to And\,XVIII.
According to their data, And\,XVIII has the heliocentric systemic velocity 
$v_{sys} = -332.1\pm2.7$ \kms and the velocity dispersion $\sigma_v = 9.7\pm2.3$ \kms.
The authors note, that the measured $v_{sys}$ is very close to M31's $v_{sys}$. They 
pointed out that at a distance of about 600 kpc from M31 And\,XVIII is near its apocenter 
and therefore at rest with respect to M31.

In this work we present new observations of the And\,XVIII dwarf galaxy 
obtained aboard the Hubble Space Telescope with the Advanced Camera for Surveys
(HST/ACS), which allow us to measure an accurate photometric distance
and the detailed star formation history of the galaxy based on the photometry
of the resolved stellar populations of And\,XVIII. These new data can
shed light on the origin and evolution of the dwarf spheroidal galaxy.

\section{Observations and photometry}
Observations of the dwarf galaxy And\,XVIII were made with the ACS/HST
on June 20, 2014 within the SNAP project 13442 (PI: R.B.Tully). 
Two images were obtained in the F606W and F814W
filters with the exposures 1100 s in each. Figure~\ref{fig:ima} shows
the F606W image of the galaxy. This low surface brightness galaxy is
very well resolved into individual stars. They can be distinguished 
in the upper chip of the ACS image. A number of bright foreground Galactic
stars contaminate the image.

We use the ACS module of the {\sc DOLPHOT} software package\footnote{http://americano.dolphinsim.com/dolphot/}
by A. Dolphin for photometry of resolved stars.
The data quality images were used to mask bad pixels. 
Only stars of good photometric quality were used in the
analysis, following the recommendations given in the {\sc DOLPHOT} User's 
Guide. We have selected the stars with signal-to-noise (S/N) of at least 
five in both filters and $\vert sharp \vert \le 0.3$.
The resulting colour-magnitude diagram of And\,XVIII is presented in Fig.~\ref{fig:cmd}.

It is well known, that artificial star tests are the most accurate way 
to estimate the photometric errors, blending and 
incompleteness in the crowded fields of nearby resolved galaxies.
These tests were performed for And\,XVIII using the same  {\sc DOLPHOT}
reduction procedures. 
A significant number of artificial stars were generated 
in the appropriate magnitude and colour range so that
the distribution of the recovered magnitudes is adequately sampled. 
Photometric errors are indicated in Fig. 1. In the CMD we show the stars 
with signal-to-noise $\ge$ 5 in both F606W and F814W filters. According to 
the artificial star experiments, the 50\% completeness level is
appearing at F814W $\simeq$ 26.7 mag and at F606W $\simeq$ 27.6 mag. 

\section{Colour-magnitude diagram and foreground contamination}

\begin{figure}
\includegraphics[height=11cm]{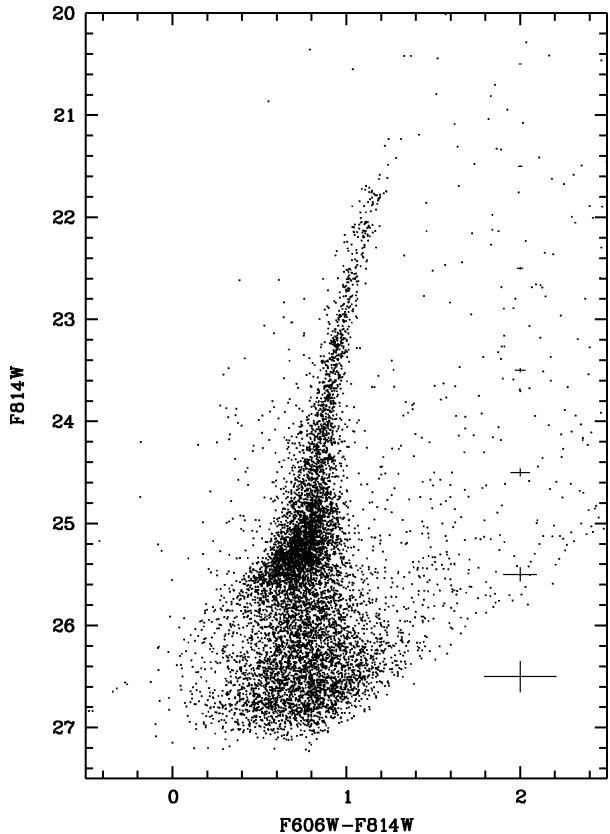}
\caption{And\,XVIII colour-magnitude diagram. 
Photometric errors are indicated by the bars at the right in the CMD.
}
\label{fig:cmd}
\end{figure}
The colour-magnitude diagram of And\,XVIII (Fig.~\ref{fig:cmd}) shows a thin
and well distinguished red giant branch (RGB). It is not heavily populated
in the upper part, and only a few probable upper AGB stars are seen above
the RGB tip at F814W = 21.7 mag. The most abundant feature in the CMD is the red clump
visible in the magnitude range between about F814W=24.6 mag and F814W=25.6 mag, 
well above the photometric limit. The blue ward stub of a horizontal branch (HB) is poorly distinguished.
It approaches the photometric limit at the blue end.
However, traces of a scattered HB population can be seen 
to the lower and bluer part of the CMD. There is no evidence of a blue main sequence, as 
is typical for dwarf spheroidal galaxies lacking ongoing star formation.

The stars in the CMD redder than about $F606W-F814W$=1.6-1.8 obviously
belong to our Milky Way galaxy (MW). To account for this contamination we need to construct
a colour-magnitude diagram of the MW in the direction of And\,XVIII. We cannot use
the HST/ACS image of And\,XVIII itself, because the galaxy occupies a significant 
part of the frame. To account for the MW contamination in this region, we
have used the \textsc{TRILEGAL} program \citep{trilegal}, that computes synthetic
colour-magnitude diagrams for the specified coordinates in the sky and given
parameters of the Milky Way model. 
Forty synthetic CMDs were constructed with \textsc{TRILEGAL} and then averaged
to avoid stochastic errors in synthetic CMDs. Random and systematic photometric
uncertainties and completeness measured from artificial star experiments were
applied to the synthetic CMDs. Resulting contamination by the MW stars was
determined to be 104 stars over the CMD of And\,XVIII. These stars were statistically
excluded from further analysis.

\section{Distance determination}
A precise knowledge of the distance is necessary to determine
the star formation history from a CMD analysis.
A photometric distance to And\,XVIII dwarf galaxy was first estimated
by \citet{mccon2008} in the framework of their CFHT/MegaPrime photometric survey of M31.
They reported the TRGB distance of And\,XVIII $D = 1.36\pm0.09$ Mpc.

The present HST/ACS photometry is deeper and it has a higher quality. There are also
a number of improvements implemented to the TRGB method itself.
We have determined the photometric TRGB distance with
our {\it trgbtool} program which uses a maximum-likelihood algorithm 
to obtain the magnitude of TRGB from the stellar
luminosity function \citep{makarov06}. 
The measured TRGB magnitude is $F814W_{TRGB} = 21.70_{-0.14}^{+0.06}$ in the ACS 
instrumental system. 
Using the calibration for the TRGB distance indicator by 
\citet{rizzietal07} and the Galactic extinction E(B-V) = 0.093 from \citet{schlafly}, 
we derived the true distance modulus for 
And\,XVIII: $25.62_{-0.17}^{+0.09}$ ($D = 1.33_{-0.09}^{+0.06}$ Mpc).
This new distance is in a good agreement with the previous estimation.

\section{Star formation history}

\subsection{The method}
The observed distribution of stars in the colour-magnitude diagram is a linear 
superposition of all the stars, at various stages of evolution, born in the galaxy 
during her life. Several additional parameters also have a strong influence
on this photometric distribution: the distance to the object,
presence of external and internal absorption and photometric errors.

The quantitative star formation and metal enrichment history of And\,XVIII was determined from 
the stellar photometry results using our StarProbe package \citep{mm04}. 
This program adjusts the observed photometric distribution of stars in the 
colour-magnitude diagram against a positive linear combination 
of synthetic diagrams.

Observation data and model data are presented in the form of Hess diagrams,
which are two-dimensional histograms
showing the number of stars in a certain range (bin) of
magnitudes and colour indices. The optimum size of the bin
depends on the characteristics of the photometric data.
On the one hand, the bin must be large enough to contain a considerable 
number of stars. On the other hand, it should 
clearly reflect the characteristics of the distribution
stars in the CMD. The size of the cells used for And~XVIII is 0.05\,mag 
in luminosity and colour.

The main and most time consuming step in determining the SFH
is to build a model CMD.
The model Hess diagrams are built on the basis of the theoretical stellar
isochrones, each corresponding to a particular age and
metallicity.  The whole set of these models covers a wide
range of ages and metallicities of stellar populations.
We used the Padova2000 set of theoretical isochrones \citep{girardi00}.
The distance is adopted from the present paper (see above) and 
the Galactic extinction is taken from \citet{schlafly}.
The accuracy and completeness of photometry is given in tabular form, obtained
as a result of the simulation of a large number of artificial stars.
We are building an analytic function of 
the distribution of stars in the Hess diagram for each isochrone of
a certain age and metallicity, taking into account the IMF,
photometry errors, the Hess diagram bin size, distance modulus and the Galactic
absorption. We use the Salpeter IMF: $\rho(m)\,dm\sim m^{-2.35}dm$.

We have taken into account the presence of unresolved binary stars (binary fraction), 
following our recipe in \citet{makarova2010}. We suppose the binary fraction to be 30
per cent, taking the mass function of individual stars and the main component 
of a binary system to be the same. 
The mass distribution for the second component was taken to be flat 
in the range 0.7 to 1.0 of the main component mass.

We interpolate the original isohrones in age to avoid discontinuities in the CMD,
so that the adjacent isochrones have maximum separation less than 0.03 mag.
We do not produce interpolation in metallicities, using the original set of metallicities.
For each isochrone we calculate the corresponding artificial Hess diagram,
which reflects the probability to find a star in a given bin.
For the star formation reconstruction we split the lifetime into relatively small periods
with 2 Gyr steps for the old stellar populations and 0.5 Gyr steps near 1 Gyr and younger.
For each time period we combine the partial artificial Hess diagrams assuming 
a constant star formation rate during this period.
As the result we construct a set of the Hess diagrams, covering range of ages from 4\,Myr to 14\,Gyr and metallicities from $Z = 0.0001$ to $Z = 0.03$.

The observed CMD is a linear combination of stars over all episodes of star formation. 
For the reconstruction of the SFH we try to find a linear positive combination of model CMDs that 
best fit the observations.
Firstly, we find the most significant episodes of star formation, which differs from 
zero with given probability, using a stepwise algorithm. 
After that we determine star formation rates using a maximum likelihood method taking into account 
that the numbers of stars in bins of a Hess diagram obey Poisson statistics. 
The probability of a given realization of the Hess diagram with $N_i$ stars in $i$-bin is
\begin{equation}
P=\displaystyle\prod_i \frac{(Xb)_i^N}{N_i!}\exp({-(Xb)_i}),
\end{equation}
where $Xb$ represents of linear combination of a stack of model CMDs with star formation rate b.
The maximum likelihood problem is equivalent to the minimization of function:
\begin{equation}
\mathcal{L}=-\log{P}\displaystyle\sum_i (Xb)_i - N_i \log (Xb)_i + \log N_i!
\end{equation}
The several thousand bins for the star formation reconstruction is large enough to use Wilks' theorem 
for confidence estimation.
It tell us that the difference between the full $\log P_{\rm F}$ and reduced $\log P_{\rm R}$ log-likelihoods follows a $\chi^2$-distribution
derived as:
\begin{equation}
-2 \left[ \log P_{\rm R} - \log P_{\rm F} \right] \sim \chi^2_{m-n},
\end{equation}
with the degrees of freedom equal to the difference of number of free parameters between full, $m$ and reduced, $n$, models.
Thus, in our case of $k$ independent episodes of star formation, a $1-alpha$ confidence region for star formation rates $b$ can be derived as:
\begin{equation}
\mathcal{L}(b)-\max_{b}\mathcal{L}\leq\frac{1}{2}C_{k,1-\alpha},
\label{e:confbounds}
\end{equation}
where $C_{k,\gamma}$ is the $\gamma$th quantile of a $\chi^2_{m}$ distribution.
Equation \ref{e:confbounds} gives us a way to estimate confidence intervals for derived values.

The resulting star formation history (SFH) is shown in 
Fig.~\ref{fig:sfh}.

The 1\,$\sigma$ errors of the estimated SFR during each period of non-zero star formation
are derived from the computed confidence levels.
The periods with insignificant star formation were excluded from the fitting process.
We assign them to be identically equal to zero. 
Thus, we do not determine the errors for the star formation with zero rates.
The confidence intervals are reliable estimation of the random errors of the SFR.
They do not include the systematic uncertainties related to specific choice of the stellar evolution models and IMF, 
as well as an influence of variations in extinction and distance estimation.
Thereby, the error bars reflect only the random errors of SFR reconstruction using Padova2000 isochrones set.
It is the lower limit of the overall uncertainties. 
It is worth to note, that the detailed analysis of systematic and random errors, which could appear in the process of maximum likelihood fitting of artificial CMDs to the real data, was performed by \citet{dolphin2012,dolphin2013}. The results confirms the importance of an isochrone set uncertainties in the SFR confidence levels. In particular, these systematic errors could dominate for the relatively shallow CMDs. However, this kind of analysis is far beyond the scope of the present article.

\subsection{SFH of And\,XVIII}
\begin{figure*}
\includegraphics[width=12cm]{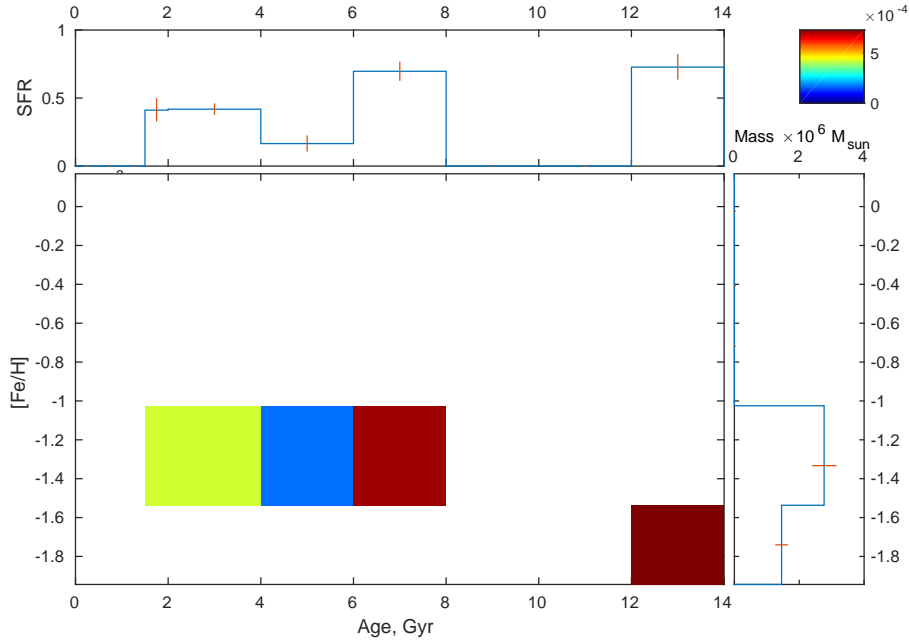}
\caption{The star formation history of And\,XVIII.
The top panel shows the star formation rate (SFR) ($M_\odot$/yr) against the
age of the stellar populations.
The bottom panel represents the metallicity of stellar content as a function of age.
The coloured boxes represent the periods of star formation for the given metallicity.
Their heights are shown only for the visualisation and does not reflects the range of metallicities.
The right panel is stellar mass vs. metallicity.
The formal errors in SFR are indicated with the vertical bars.
}
\label{fig:sfh}
\end{figure*}

\begin{figure*}
\includegraphics[width=12cm,clip]{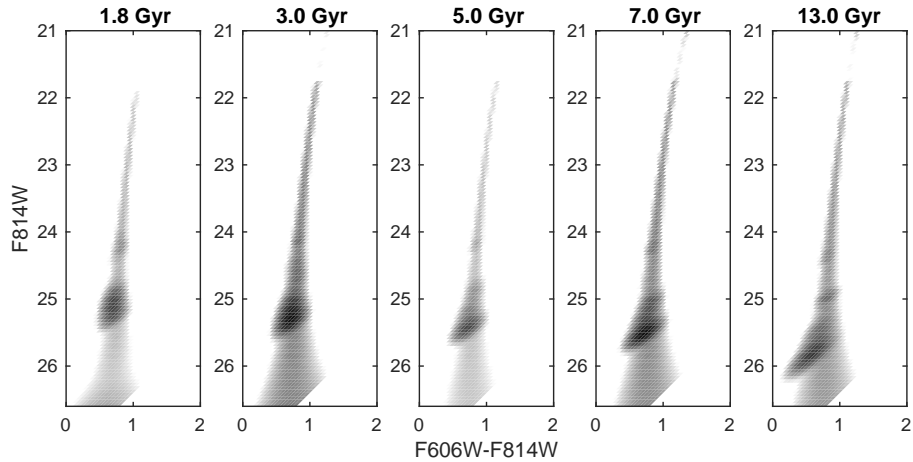}
\caption{Andromeda\,XVIII star formation history reconstruction.
The panels correspond to the five episodes of star formation. The grayscale represents the predicted 
number of stars in each episode. The most intensive episodes are those at 3 and 7 Gyr, which 
mainly describe the big and stretched red clump in the real CMD. }
\label{fig:reconstruct}
\end{figure*}

\begin{figure*}
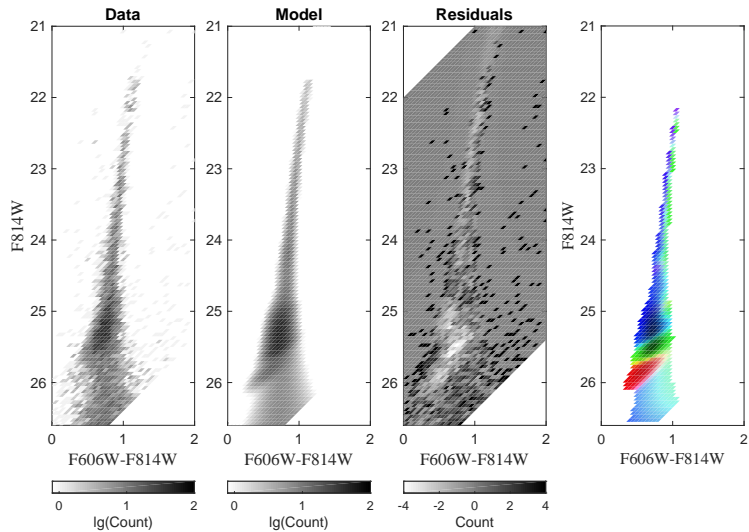

\includegraphics[height=7cm,clip]{fig5_1.eps}
\raisebox{8.5mm}{\includegraphics[height=6.15cm,clip]{fig5_2.eps}}
\caption{
The data panel shows the observed Hess diagram of Andromeda\,XVIII.
The model panel presents the reconstructed Hess diagram.
The number of stars per bin is coded by grey colour on a logarithmic scale in both the data 
and the model panels.
The residual panel shows the difference between the observed and model diagrams. The values in each bin
were normalized by the square root of the respective model value, which provides an estimation 
of the Poisson noise.
Grey colours vary from white for negative to black for positive residuals.
Even the most discrepant regions are within $\pm 5 \sigma$ deviation.
In the colour version of the model CMD the old 13 Gyr population is red,
the 5+7 Gyr populations are green and the 1.5+3 Gyr populations are blue.}
\label{fig:reconstruct2}
\end{figure*}

Figures~\ref{fig:sfh}, \ref{fig:reconstruct} and \ref{fig:reconstruct2} demonstrate the star 
formation rate and 
metallicity changes during the life of And\,XVIII. According to our calculations, 
an ancient burst of star formation occurred 12--14 Gyr ago. The mean SFR
during this period was $7.3\pm0.9\times$10$^{-4}\,M_{\odot}/yr$. The metallicity 
of the stars that were formed were predominantly in the interval [Fe/H]=$-2$ -- $-1.6$ dex. 
The CMD indicates that there followed a quiescent
period in the star formation of And\,XVIII from 8 to 12 Gyr ago.
This phenomenon is widespread in dwarf galaxies. It is known that in such objects
the star formation process is complex, and bursts alternate with periods 
of lower rates of star formation (\citet{weisz2011}, \citet{savino2015}, \citet{karachentsev1999}, and references therein).

However, at the And\,XVIII distance of 1.33 Mpc we have reached an absolute I magnitude
about +1 in a single HST/ACS orbit, i.e. the fainter stellar populations
like the horizontal branch and the lower part of the main sequence
are below our photometric limit.
Without the information revealed in this part of a CMD it is difficult to resolve 
the age-metallicity-SFR relation for the oldest ( $>6 - 8$ Gyr) star formation
events, due to tight packing of the corresponding isochrones at the 
brightest part of the CMD.
Nevertheless, \citet{weisz2011} have demonstrated that a recovered SFH does not depend on
the photometric depth, but rather on the number of stars in the CMD, if the using stellar models are 
known exactly. But the authors also note, that stellar models are not always self-consistent,
and the measured SFR may be systematically shifted into a particular time bin, depending on 
the stellar models and photometric depth of the CMD. However, in the case of And\,XVIII
the detailed analysis of the recovery of the red clump (see below in this Section) supports each of 
the SFH claims.

Our measurements also indicate a process of continuous star formation from 1.5 to 8 Gyr ago.
These stars must have higher metallicity at [Fe/H]=$-1.6$ -- $-1.0$ dex. 
We see evidence from the synthetic diagram of a variation of
the mean intensity of star formation from 1.6 to $7\times$10$^{-4}\,M_{\odot}/yr$.
There is no sign of recent/ongoing star formation in the last 1.5 Gyr of And\,XVIII's life.
The mass fraction of the ancient and intermediate age stars is 34 and 66 per cent, respectively. 
The total measured stellar mass of And\,XVIII is $4.2\times$10$^6\,M_{\odot}$.
In Fig.~\ref{fig:reconstruct} we show Hess diagrams of the reconstructed stellar populations. 
The panels correspond to the five episodes of star formation. The grayscale is calibrated to the predicted 
number of stars in each episode. The most intensive are the 3 and 7 Gyr episodes, that describe 
the big and stretched red clump in the observed CMD.
In Fig.~\ref{fig:reconstruct2} the observed and combined model CMDs are shown.
The number of stars per bin is coded by grey colour on a logarithmic scale in both the data and the model panels.
The residual panel shows the difference between the observed and model diagrams. The values in each bin
were normalized by the square root of the respective model value, which provides an estimation 
of the Poisson noise.
Grey colours vary from white for negative to black for positive residuals.
Even the most discrepant regions are within $\pm 5 \sigma$ deviation.
The very weak AGB is fitted quite well, with $8\pm3$ observed stars against $16.6\pm4.1$ modelled.

\citet{kirby} have measured the metallicity of a few red giant stars in And\,XVIII from individual
stellar spectra. They calculate the value of [Fe/H] = $-1.35\pm0.20$, which is in good
agreement with our measurements of the metallicity of red giants of age between 1.5 and 8 Gyr.
Observed and model parameters for the And\,XVIII galaxy are gathered in Table 1.

We summarize the salient features described by our age-metallicity model.  
There cannot have been significant star formation in the last 1.5~Gyr because of the absence of obvious markers of youth.
The stunted horizontal branch tells us that there was an ancient episode of star formation ($\sim 13$ Gyr), though limited.
The prominence and extended luminosity range of the red clump requires that there are intermediate age stars.
The fainter red clump stars imply older ages, up to 7~Gyr while the brighter red clump stars are compatible with ages as young as 1.5~Gyr.
The plot with colours in Fig.~\ref{fig:reconstruct2} emphasizes the evidence for these claims.  
The oldest population, in red, manifests itself mainly in the HB.  
Stars at $5-8$ Gyr, in green, have built the fainter range of the red clump. 
Stars in the range $1.5-3$ Gyr produce a more luminous red clump.  
The upper part of the red giant branch is thin. There is a formal separation between the younger intermediate age stars in blue and the older intermediate age stars in green because the same metallicity is taken for both populations.
If the younger stars have slightly higher metallicity then the populations would merge along the bright extension of the RGB.
The ancient population is buried within the upper RGB by happenstance with its lower metallicity. 

\begin{table*}
\centering
\caption{General parameters of And\,XVIII}.
\label{table:general}
\begin{tabular}{lll}
\hline
                                        & And\,XVIII                      & Source   \\ 
\hline
Position (J2000)                        & 00h02m14.5s$+$45d05m20s         & NED      \\
$E(B-V)$, mag                           & 0.093                           & \citet{schlafly}  \\
$V_T$, mag                              & 15.50$\pm$0.24$^a$              & this work  \\
$I_T$, mag                              & 14.68$\pm$0.15                  & this work  \\
$M_V$, mag                              & $-10.41\pm0.28$                 & this work  \\
$M_I$, mag                              & $-11.10\pm0.20$                 & this work   \\
Central surface brightness in $V$, mag arcsec$^{-2}$  &  23.96$\pm$0.02   & this work  \\
Central surface brightness in $I$, mag arcsec$^{-2}$  &  23.14$\pm$0.02   & this work  \\
Exponential scale length in $V$, arcsec &  $26.2\pm0.4$                   & this work  \\
Exponential scale length in $I$, arcsec &  $26.8\pm0.4$                   & this work  \\       
Axial ratio $b/a$                       & 0.99                            & LV database$^b$    \\
Heliocentric velocity $v_{sys}$, \kms{} & $-$332.1$\pm$2.7                & \citet{tollerud}  \\
Total velocity dispersion, $\sigma$ \kms{} & 9.7$\pm$2.3                  & \citet{tollerud}  \\
Distance modulus, mag                   & $25.62_{-0.17}^{+0.09}$         & this work  \\
Distance, Mpc                           & $1.33_{-0.09}^{+0.06}$          & this work  \\
Linear distance to M~31, kpc            & $579\pm87$                      & this work  \\
Mass fraction of oldest stars (12--14 Gyr) & $34\pm5$ \%                  & this work  \\
Mean metallicity of oldest stars, [Fe/H], dex& $-1.74\pm0.20$             & this work  \\
Mean SFR 12--14 Gyr ago, $M_\odot$/yr   & $7.3\pm0.9\times$10$^{-4}$      & this work  \\
Fraction of intermediate age stars (1.5--8 Gyr)   & $66\pm6$ \%           & this work  \\
Mean metallicity of intermediate age stars, [Fe/H], dex& $-1.33\pm0.20$   & this work  \\
Total stellar mass, M$_\odot$           & $4.2\pm0.3\times$10$^6$               & this work  \\
\hline
\multicolumn{3}{l}{\footnotesize{$^a$The total magnitudes are not corrected for Galactic extinction}} \\
\multicolumn{3}{l}{\footnotesize{$^b$http://www.sao.ru/lv/lvgdb/}} \\
\hline
\end{tabular}
\end{table*}

\subsection{Star formation in nearby isolated dwarf galaxies}
Table 2 provides the star formation parameters for 6 dwarf galaxies at the periphery of 
the Local Group. The star formation histories of dSphs KKR25,
KKs03 and And\,XVIII, as well as dTr KK258 were uniformly measured by us from 
HST/ACS observations. The nearer dwarf irregulars DDO210 and Leo A were analysed by \citet{cole}
from HST/ACS photometry extending about 2 mag deeper to include fainter main sequence stars.
The CMDs of the two latter objects resolve the earliest epoch of star 
formation in some detail.

All the galaxies are situated well beyond the individual virial radii of M31 and the Milky Way and the two dSphs
KKR25 and KKs03 are extremely isolated. It can be expected that the evolution of all
these objects have been unfettered by external tidal influences. The measured total stellar masses are low for all 
the galaxies, not exceeding about $2\times$10$^7$ M$_\odot$. However,  the ACS fields do not include faint outskirts 
so stellar mass estimates are somewhat truncated.

In this small sample, the systems with the largest stellar masses, those above $10^7$~M$_{\odot}$, 
had the largest fraction of their stars formed early.  The two irregulars started very slowly with star formation.
And\,XVIII is intermediate.

One can speculate that early evolution of very isolated dwarf galaxies 
is controlled mostly by the value of the potential well. For deeper wells larger amount of 
gas accretes faster towards the centre, cools down and the bulk of stars forms in the shortest time.

\begin{table*}
\centering
\caption{Early star formation in the local isolated dwarfs}
\label{table:general}
\begin{tabular}{lllcccll}
\hline
Name   &  Type  &  Distance$^a$  & $M_{\ge12 Gyr}^*$ & $M_{\ge8 Gyr}^*$ & $M_T^*$ & SFR $\ge$ 12 Gyr & Source \\
       &        &    Mpc     &    \%       & \%     & M$_\odot$ &  $M_\odot$/yr &    \\
KKR25  &   dSph &  $1.93\pm0.07$ & 62 & 62 & $3.0\times10^6$ & $1.7\pm0.2\times10^{-3}$ & \citet{mak12} \\
KK258  &   dTr  &  $0.84\pm0.09$ & 70 & 70 & $2.2\times10^7$ & $7.9_{-3.2}^{+4.8}\times10^{-3}$ & 
\citet{kar14} \\
KKs03  &   dSph &  $2.12\pm0.07$ & 74 & 74 & $2.3\times10^7$ & $8.7\pm0.4\times10^{-3}$ & \citet{kar15} \\
DDO210 &   dIrr &  $0.97\pm0.06$$^b$ & 10 & 24 & $4.0\times10^6$ & $2.2\pm1.0\times10^{-4}$ & \citet{cole} \\       
LeoA   &   dIrr &  $0.75\pm0.16$$^b$ &  4 & 11 & $3.5\times10^6$ & $9.0_{-6.4}^{+2.2}\times10^{-5}$ & \citet{cole} \\ 
AndXVIII  &   dSph &  $0.58\pm0.09$ & 34 & 34 & $4.2\times10^6$ & $7.3\pm0.9\times10^{-4}$ & this work \\
\hline
\multicolumn{8}{l}{\footnotesize{$^a$Distance from nearest giant galaxy}} \\
\multicolumn{8}{l}{\footnotesize{$^b$Distance value from \citet{jacobs09}}} \\
\hline
\end{tabular}
\end{table*}

\section{Total and surface photometry}
Total and surface photometry of And\,XVIII was made with fully processed 
distortion-corrected HST/ACS F606W and F814W images.
Background stars were removed from the frames by fitting of a first order 
surface in a rectangular pixel-area in the nearest neighbourhood of a star. 
The sky background in the ACS images is insignificant but, to remove
possible slight large scale variations, 
the sky was approximated by a tilted plane, created from 
a two-dimension polynomial, using the least-squares method. The accuracy 
of the sky background determination is about 1 -- 2 \% of the original sky
level.

To measure total galaxy magnitude in each band, the galaxy image was at first
fitted with concentric ellipses. Then integrated
photometry was performed in concentric ellipses with the defined parameters
from a centre to the faint outskirts of the galaxy.
The total magnitude was then estimated as the asymptotic value of
the obtained radial growth curve. The measured total magnitudes are 
$V = 15.50\pm0.24$ mag and $I = 14.68\pm0.15$ mag. The estimated errors 
include the photometry and sky background uncertainties, as well as
the transformation errors from instrumental ACS magnitudes to the standard
V and I magnitudes \citep{sirianni}. The corresponding absolute magnitudes
of And\,XVIII are $M_V = -10.41$ and $M_I = -11.10$, taking into account
Galactic extinction \citep{schlafly}, and the distance modulus from the
present paper (see above).

\begin{figure}
\includegraphics[width=8cm]{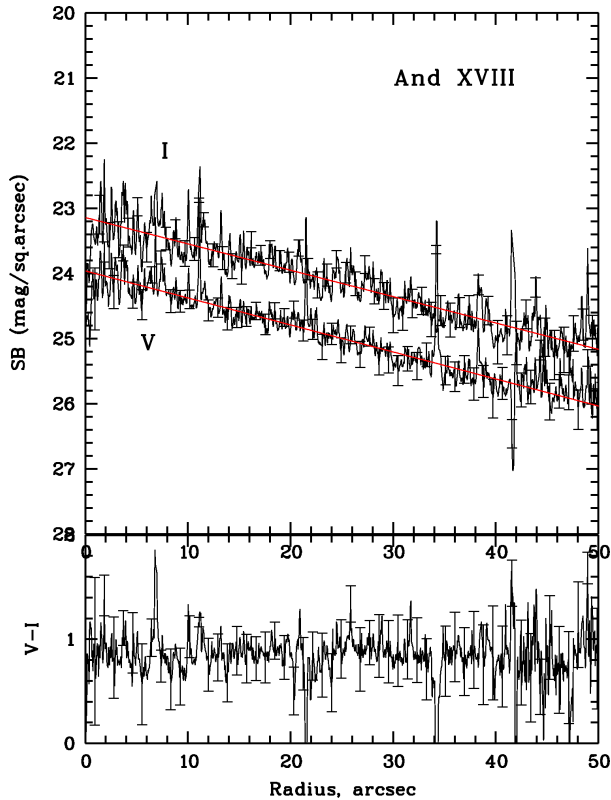}
\caption{
The surface brightness profiles of And\,XVIII in V and I bands (top panel).
The colour profile in shown in the bottom panel. Photometric uncertainties
are indicated by the vertical bars.
}
\label{fig:profile}
\end{figure}

Azimuthally averaged surface brightness profiles for And\,XVIII
were obtained by differentiating the galaxy growth curves with
respect to semiaxes. The resulting profiles in V and I bands
are displayed in Fig.~\ref{fig:profile}. 
The very low surface brightness of this dSph galaxy results in profiles that are quite 
noisy and shallow.

It is well known that surface brightness profiles of dwarf galaxies
(both irregular and spheroidal) and also disks of spiral galaxies can
be fitted by an exponential intensity law of brightness distribution \citep{vau}
$$ I(r) = I_0*exp(-r/h)$$
or, in magnitudes per sq.sec
$$ \mu(r) = \mu_0+1.086*(r/h), $$
where $\mu_0$ is the central surface brightness and $h$ is the
exponential scale length.
The surface brightness profiles of And\,XVIII galaxy are well-fitted by 
an exponential law. The unweighted exponential fits to the surface brightness 
profiles were obtained by linear regression. The derived central surface brightness
is $\mu_0^V = 23.96\pm0.02$ mag\,arcsec$^{-2}$ and $\mu_0^I = 23.14\pm0.02$ mag\,arcsec$^{-2}$. 
The uncertainties are formal fitting errors. The exponential scale lengths
are $h_V = 26.2$ arcsec and $h_I = 26.8$ arcsec, 170 and 174 pc respectively.

\section{The Local Group structure and isolated dSphs}
\begin{figure*}
\includegraphics[width=18cm]{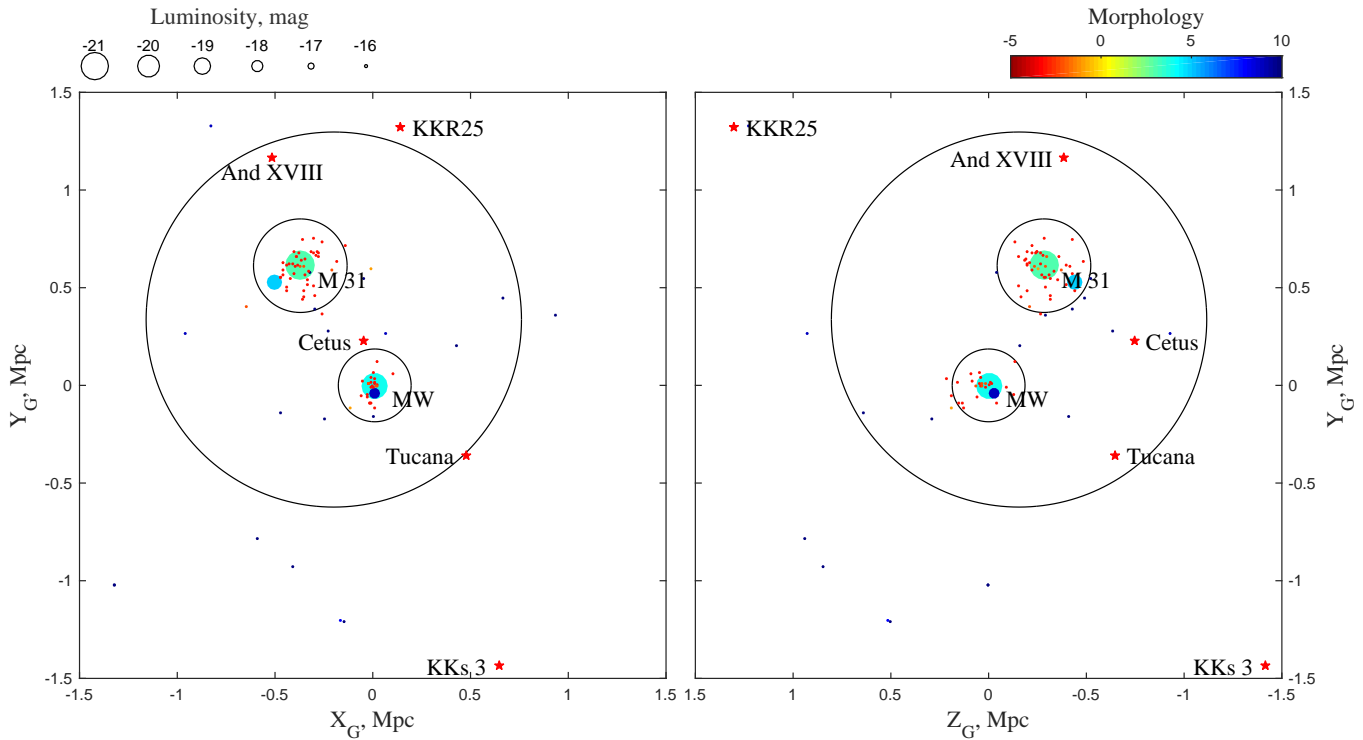}
\caption{A panorama of the Local Group in the galactic coordinates.
The figure shows the projection of nearby galaxies in a cube $\pm 1.5$~Mpc from us.
The left panel is a projection on the galactic plane XY, while the right 
panel is the ZY view of the distribution of galaxies. The colour of a dot represents 
the morphology of the galaxy according to the colour bar.
The size of a galaxy corresponds to its luminosity as shown in the legend panel.
The five most isolated dSphs galaxies are represented by red stars.
The big circle encloses the Local Group shows the sphere of zero-velocity with 
$R_0=0.96$~Mpc \citep{Karachentsev+2009}.
The small circles around Milky Way and M~31 are virial radii, $R_{200}$, 
that correspond to masses $0.8\times10^{12}$ and 
$1.7\times10^{12}$ $M_\odot$ respectively \citep{Diaz+2014}.}
\label{fig:structure}
\end{figure*}

The structure of the two subgroups of the Local Group is demonstrated in 
the Fig.~\ref{fig:structure}. One subgroup is concentrated around our Milky Way 
galaxy and the other around Andromeda galaxy (M31). And\,XVIII is situated
quite far from both giant spirals. It is associated with M31, being
situated at the distance of 579 kpc from this gravitational centre, but 
at roughly twice the virial radius of the M31 group.
\citet{mccon2008} note that And\,XVIII lies outside the known satellite planes 
around M31.

In Fig.~\ref{fig:structure} there are clearly visible concentrations of dwarf spheroidal 
galaxies, the satellites of MW and M31 within their virial radii. This phenomenon is well 
known as the distance-morphology relation
 \citep{grebel2003, grebel2005}. Although each family 
of the giant spirals satellites is quite extensive, exceptions from the familiar type dependence 
is rare. We have marked such objects with red stars: Cetus and And\,XVIII are located within 
the zero-velocity sphere of the Local Group, while Tucana, and especially KKR25 and KKs3, 
rightfully can be identified as isolated dwarf spheroidal galaxies.
Detection of these dwarfs is of great importance for the understanding of star formation 
and evolution of dwarf galaxies.

In the work of \citet{gp2009} the authors note that most of the satellite dwarfs of the
Milky Way and Andromeda situated within 260 - 280 kpc are not detected
in HI, and this fact could be an indication that the predominant
mechanism of gas loss is associated with the proximity to the main
galaxy. There are several models
that attempt to explain the origin of dSph galaxies by considering
different mechanisms. Ram-pressure stripping, tidal stripping,
feedback from supernovae or stellar winds, and the effects of
reionization were considered as possible mechanisms of gas loss. 
Models based on tidal and ram-pressure
stripping reasonably explain the formation of dSphs 
\citep{gp2009, assmann2013}. In these models, the dSph galaxies are
formed due to the interaction between a rotationally supported dwarf
irregular galaxy and a MW-sized host galaxy. Thus, the fact that we
rarely find isolated dSph galaxies indicates the relevance
of the proposed tidal and ram pressure mechanisms.

Nevertheless, the rather isolated dSphs at the far periphery of 
the Local group indicate that other mechanisms of evolution 
should be considered.
\citet{teyssier} used the Via Lactea II cosmological simulations to
distinguish between dwarf galaxies within the Local Group that may have 
passed through the virial volume of the Milky Way, and those that have not.
According to their modelling, Tucana and Cetus may have passed through 
the Milky Way in the past. In these cases 
recent star formation quenching, star burst episodes and low gas fraction 
could be explained by an encounter with the MW.

And\,XVIII dwarf spheroidal galaxy is situated at the linear distance of 
579 kpc from the Andromeda galaxy. According to \citet{tollerud}, the galaxy
has the systemic velocity $v_{sys} = -332.1\pm2.7$ \kms, whereas M31 has
mean heliocentric radial velocity $-290\pm16$ \kms.
It is a question whether this galaxy experienced any interaction in the past.
In the paper of \citet{watkins} the orbital properties of the M31 satellites are estimated 
by a statistical analysis using a combination of the timing argument and 
phase-space distribution functions. This analysis provides periods of the known dwarf 
satellite galaxies, allowing us to distinguish the galaxies likely in the first infall
to the M31 group. And\,XVIII has the period T=$19.5\pm1.0$ Gyr. This period 
makes this dwarf galaxy a first-approaching satellite. 
Unfortunately,  the authors acknowledge 
that And XVIII may not be well described by their models due to the large separation distance.
Anyway, it is highly possible that star formation cessation 
in this dwarf spheroidal galaxy has occurred without the influence of an interaction with the giant spiral M31.

According to our measurements (see description in section 5), And\,XVIII had an extended
period of star formation from 1.5 to 8 Gyr ago, a distinct star formation
episode from 12 to 14 Gyr ago, and no signs of star formation in the last 1.5 Gyr.
Taking into account the studies mentioned above, we suggest, that the gas loss and
star formation quenching in the last 1.5 Gyr were driven by internal processes in 
the galaxy itself rather than by any external influence. 

\section{Concluding remarks}
We present new observations with HST/ACS of the Andromeda\,XVIII dwarf galaxy situated in 
the Local Group of galaxies and associated with giant spiral M\,31. We analyse the star formation
history and possible evolution of the galaxy and summarize the results of
our study as follows:

\begin{itemize}

\item We obtained a colour-magnitude diagram of the And\,XVIII as a result of 
the precise stellar photometry of the resolved stars possible with deep F606W (V) and F814W (I) HST/ACS
images. The colour-magnitude diagram of And\,XVIII reveals a thin
and well distinguished red giant branch. The most abundant feature in the CMD is the red clump,
visible well above the photometric limit. There is no pronounced blue main sequence, a characteristic of
dwarf spheroidal galaxies that lack ongoing star formation.

\item The high quality stellar photometry allows us to derive an accurate distance to And\,XVIII
using the tip of the red giant branch distance indicator. The measured TRGB magnitude 
is $F814W_{TRGB} = 21.70_{-0.14}^{+0.06}$ in the ACS instrumental system. 
Using the calibration for the TRGB distance indicator by 
\citet{rizzietal07} and the Galactic extinction E(B-V) = 0.093 from \citet{schlafly}, 
we derived the true distance modulus for 
And\,XVIII of $25.62_{-0.17}^{+0.09}$ ($D = 1.33_{-0.09}^{+0.06}$ Mpc).
Taking into account this new distance,
And\,XVIII dwarf spheroidal galaxy is 
situated at the linear distance of 579 kpc from M31, well
outside of the virial radius of the M31 halo.

\item We performed total and surface photometry of And\,XVIII with the fully processed 
distortion-corrected HST/ACS F606W and F814W images. 
The measured total magnitudes are 
$V = 15.50\pm0.24$ mag and $I = 14.68\pm0.15$ mag. The correspondent absolute magnitudes
of And\,XVIII are $M_V = -10.41$ and $M_I = -11.10$, taking into account
Galactic extinction \citep{schlafly} and the distance modulus given directly above.
The surface brightness profiles of And\,XVIII galaxy are well-fitted by 
an exponential law. The derived central surface brightness
is $\mu_0^V = 23.96\pm0.02$ mag\,arcsec$^{-2}$ and $\mu_0^I = 23.14\pm0.02$ mag\,arcsec$^{-2}$. 
The uncertainties are formal fitting errors. The exponential scale length
is $h_V = 26.2$ arcsec (170 pc) and $h_I = 26.8$ arcsec (174 pc).

\item The quantitative star formation and metal enrichment history of And\,XVIII was determined from 
the stellar photometry results using our StarProbe package \citep{mm04}. According to our calculations, 
an ancient burst of star formation occurred 12--14 Gyr ago. The metallicity 
of the stars formed at that time is found to be in the interval [Fe/H]=$-2$ to $-1.6$ dex. 
Our model is consistent with a subsequent quiescent period in the star formation history of And\,XVIII from 8 to 12 Gyr ago.
Following that period, our measurements indicate that there was star formation from 8 to 1.5 Gyr ago, particularly prominently at the earliest and latest of those times.
The stars from this period have higher metallicity of [Fe/H]=$-1.6$ to $-1.0$ dex. 
There is no signs of recent/ongoing star formation in the last 1.5 Gyr. 
The mass fractions of the ancient and intermediate age stars are 34 and 66 per cent, respectively. 
The total stellar mass of And\,XVIII is $4.2\times$10$^6\,M_{\odot}$.

\item Taking into account this detailed star formation history of And\,XVIII, as well as 
the models in the works of \citet{teyssier} and \citet{watkins}, we compare the possible
evolution scenarios of And\,XVIII and 5 other isolated dwarf galaxies in the vicinity
of the Local Group.  It is suggested that, in isolated galaxies, gas loss and
star formation quenching, the evident current condition in dSph systems, were driven by internal 
processes in 
the galaxy itself rather than by an external influence. It is likely
that early evolution of very isolated dwarf galaxies 
is controlled mostly by the value of the potential well. For deeper wells a larger amount of 
gas accretes faster towards the centre, cools down, and the bulk of stars form in a short time.

\end{itemize}

\section*{Acknowledgements}
This research is supported by award GO-13442 from the Space Telescope Science Institute
for the analysis of observations with Hubble Space Telescope.
This study is supported by the Russian Science Foundation (grant 14--12--00965).
SFH measurement was made under the partial support from Research 
Program OFN-17 of the Division of Physics, Russian Academy of Sciences.
\bibliographystyle{mn2e}
\bibliography{and18text}   

\bsp
\label{lastpage}

\end{document}